# Training physical matter to matter

Heinrich M. Jaeger, Arvind Murugan, Sidney R. Nagel
The James Franck Institute and Department of Physics, The University of Chicago
929 E 57th St., Chicago, Illinois 60637, USA

Corresponding author: Heinrich Jaeger, h-jaeger@uchicago.edu

**Abstract**
Biological systems offer a great many examples of how sophisticated, highly adapted behavior can emerge from training. Here we discuss how training might be used to impart similarly adaptive properties in physical matter. As a special form of materials processing, training differs in important ways from standard approaches of obtaining sought after material properties. In particular, rather than designing or programming the local configurations and interactions of constituents, training uses externally applied stimuli to evolve material properties. This makes it possible to obtain different functionalities from the same starting material (pluripotency). Furthermore, training evolves a material in-situ or under conditions similar to those during the intended use; thus, material performance can improve rather than degrade over time. We discuss requirements for trainability, outline recently developed training strategies for creating soft materials with multiple, targeted and adaptable functionalities, and provide examples where the concept of training has been applied to materials on length scales from the molecular to the macroscopic.

Memorizing and forgetting, teaching and learning, improving endurance with exercise: all are processes that relate to our daily experience. These processes may seem at first to be essentially biological activities. However, by taking appropriate cues from biology one can see how similar manipulation can be applied not just to biological but also to physical matter. As an example, machine learning, the broad endeavor to teach a system to classify objects into separate meaningful categories, is in part an attempt to apply training to computer algorithms. This leads us to ask more generally: How can the broad concept of training be applied to produce desired functionality in materials?

In biological systems training provides a path toward improved performance. Moreover, biological systems can often be trained for one property and then, when no longer useful to the organism, be retrained for an entirely different purpose. Indeed, one can think of the long evolutionary history of living organisms as training to make biological systems exquisitely adaptable to new requirements from their environment. This occurs through an intricate interplay of evolution and adaptation.

For example, individuals can alter their body's musculature by targeted personal-training regimens. By lifting weights, they can become stronger in arms or legs or core; by running they can develop flexibility, endurance, and speed; by practicing a musical instrument they can train the smaller muscles to operate repetitively and smoothly to execute delicate maneuvers. Some of this training presumably occurs in the brain, but much of it also occurs in the muscles themselves – that is, by altering the size, shape and/or function of the biological material. A second, less obvious, example is Wolff's law [1] for how bone becomes stronger and more resilient due to repeated exposure to stresses; when small breaks occur, the body rebuilds those particular spots to become stronger. Exercise – *training* – evolves the material of the bone to become more useful for the specific tasks encountered and which are required of it in the course of living. The material gains this resilience by being trained for it by repetition.

As a community, we have begun to realize that physical matter – that is materials not associated with living or biological tissue – can also be trained in non-trivial ways. We are all familiar with the processing of materials. While training can certainly be viewed as a form of material processing, it occupies a rather special niche since the philosophy underlying training differs from that of typical processing protocols. As an example consider steel blades, which can be made stronger and sharper by heat tempering of the alloys [2]. In that case, the material acquires its enhanced function via exposure to temperatures or stresses far exceeding those it encounters during its intended application. Later during use it does not become stronger or sharper over time. It also cannot be *retrained* after deployment in the field to give it different material properties. An ideal trainable material, on the other hand, acquires its enhanced functionality *in-situ* under conditions similar to those during actual use. This implies that it will perform better the more it is used. Moreover, to obtain new functionality it can be retrained in the field after deployment.

The idea of training to create or enhance function is also different from the traditional philosophy underlying materials design. Conventional approaches focus on generating desired material properties by designing specific structural configurations and associated local interactions among the constituent components of a material, often at molecular scales [3-6]. Changing the targeted properties then requires careful re-programming of those local interactions. Furthermore, once identified, the associated design parameters are typically intended to remain fixed, so as to maintain a material's properties. In other words, the key goal is to find local interactions that correspond to deep local minima in the free energy.

By contrast, trainability requires that adaptive changes in those parameters must be possible and evolution of the original interactions becomes the central feature of the training process. Furthermore, during training, stimuli are applied to the material as a whole and these stimuli then find their way selectively to those local constituents whose adaptive response generates the desired overall performance enhancement. Effectively, while typical design involves deliberate manipulation of local interactions, training applies only global cues. Under an appropriate training protocol, the material reconfigures as it learns and memorizes the relevant interaction parameters on its own and without detailed control at the local level. Conceptually, this stimuli-induced but otherwise autonomous evolution of local interactions represents a radical departure from standard design approaches (**Fig. 1**).

It is exciting to envision training as a new paradigm for creating pluripotent function. Thus we could ask, taking another page from the biology textbooks, whether one can make a material platform that mimics the capabilities of stem cells so that exposure to the environment (training) can dictate the subsequent functionality of the material *after it has been deployed.* As is possible with a stem cell, the same material could subsequently be re-trained for many different purposes.

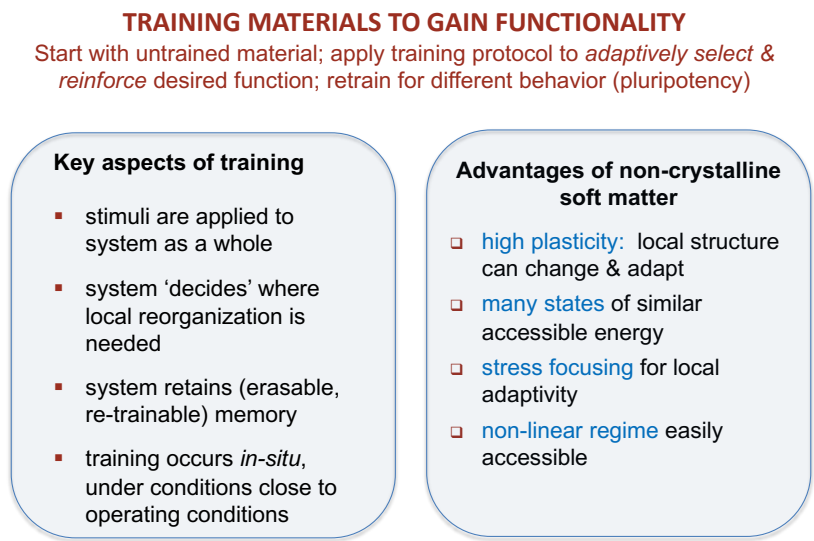

**Fig. 1.** Training as new paradigm for generating desired functionality in materials

Clearly, not all materials can be trained to produce a useful outcome. Thus, it is important to understand what is required in a material for it to be 'trainable' and potentially 'retrainable,' and what are the possibilities and limits of material training. Furthermore, for a given trainable material, a key task will be the design of appropriate training environments and regimens. Thus, both proper choice of the material platform and proper design of the training protocol will be critical for reaching a desired functionality or level of performance.

At this stage it is useful to give some specific examples of how a physical material can be trained to create behavior not normally found. One example is to exploit material aging for this purpose [7]. Aging occurs in systems that have been forced out of equilibrium either by dropping the temperature or exerting external forces. As the system continually searches the available phase space, it discovers lower free-energy configurations. The longer it explores, the lower the energy can get. We are all too familiar that this often leads to detrimental degradation of the material. However, a system can sometimes be coaxed to go downhill in energy towards a state that represents a *preferred* outcome. This state is selected by appropriately applied biases, *e.g.*, stresses or strains, on the material as a whole. In other words, such aging is *directed* to evolve in a desired manner. The inhomogeneous local response of the material allows the material to evolve because stressed regions relax at different rates than unstressed regions. In this example, the aging behaves in a greedy fashion, analogous to how a greedy algorithm works in a simulation; all that it needs to know is which direction in phase space the energy will decrease the fastest. It chooses its own pathway just by being exposed to the applied stresses or strains. Moreover, this training protocol is not relegated to the linear response of a material; it can readily be extended to non-linear properties [8]. One way of thinking about this is to consider each of the material's constituent building blocks that can age as a transient degree of freedom in the system [9]. While the material ages, these degrees of freedom change their values until they have reached a particularly low-stress regime (or until the temperature is lowered so that relaxation is slowed).

Directed aging has been used to create materials which have uncommon properties. An example is a material with a negative Poisson's ratio, $\nu$. Nearly all natural materials have a positive Poisson's ratio, $\nu$ [10]. (When stretched along one axis, a material with $\nu > 0$ becomes compressed along the axes perpendicular to it.) That is, natural materials tend to be somewhat incompressible. But elasticity theory allows both negative as well as positive Poisson's ratios. Simply by placing a material under uniform compression and then aging it in that configuration for a substantial period of time, an otherwise normal elastic material can fundamentally alter its behavior by developing a negative Poisson's ratio, known as auxetic behavior [7, 11]. Different values of $\nu$ can be achieved by varying the training conditions. Ideas like directed aging have also been used to train origami structures with multiple folding patterns to fold along specific pathways [12] in experiments.

Directed aging can be extended to *meta-properties* such as making a material *adaptable* to specific changes cued from the environment [13]. In this case, the objective is not to train for any single functionality, but rather to train the material such that it can most easily switch between a set of targeted functionalities that are incompatible. This, too, takes its inspiration from biological evolution which has led to organisms that easily adapt to sudden drastic changes in the environment [14, 15]. It turns out that by applying a training protocol that periodically switches between desired training targets, one can teach (at least in a computer model) a material to perform incompatible functionalities with minimal changes in parameters.

Another recent approach uses material training similar to directed aging that alternates between softening and stiffening of bonds (a `thumbs up/thumbs down' rule [16, 17]) to accomplish more complex tasks. Other, more involved training variants based on Equilibrium Propagation [18], such as coupled learning [19], have mathematically provable convergence. However, this benefit comes at the expense of requiring two or more copies of the material whose response to changing boundary conditions can be compared. Unlike directed aging or thumbs up/thumbs down extensions, Equilibrium-Propagation-based rules may not be a natural path that many systems automatically follow on their own. However, it has been shown that specially designed electrical circuits can be trained this way to perform a simple classification task as well as linear regression [20]. One can conceive other extensions of this approach, which can also be implemented in mechanical metamaterials, such as exploiting non-equilibrium memory [21].

In thinking about the requirements for training it is apparent that some materials can be trained more easily than others. Soft materials [22] appear particularly well suited to training as they often have a multitude of easily accessible, energetically similar configurations that training can select so as to amplify a desired property (see **Fig. 1**). In soft materials, training protocols based on applying mechanical stress provide straightforward access to *nonlinear* regimes which can help to imprint a memory [23] by triggering long-lasting changes such as plastic structural deformation [8]. However, training need not be relegated only to soft matter or to mechanical responses to applied stress or strain. It can involve any other type of response, be it electrical, magnetic, or optical. The ideas that undergird the notion of training are far from specific. Instead, training is a concept for how to create desired function. It should be applicable at different length scales from molecular to architectural and to different energy scales to include materials of conventional hard matter.

Thus, in considering different types of materials as platforms for trainable, potentially pluripotent behaviors we are looking for at least two key ingredients: first, a number of distinct responses (states) that each can retain a memory of the training outcome; second, the ability of the material to reconfigure and evolve from one state to another, meaning that these states must not only exist but be accessible during the training.

It makes sense that some degree of disorder, either in the local interactions among the components of a material or in their structural configuration, is advantageous in enabling reconfigurability [24-26]. By contrast, perfect crystals, while exhibiting exceedingly long-term configurational memory, cannot adaptively evolve their structure and thus are not trainable. Importantly, as we saw with directed aging, disorder also makes it possible for stimuli, which during training are applied on the outside of a material, to be directed to those parts inside the material that benefit the most from adaptation. Such internal focusing of uniformly applied external stimuli is well known from the physics of disordered, heterogeneous materials. For example, mechanical stress propagates through such materials along a network of paths that concentrate force on a limited number of internal spots [27, 28]. If these spots can be trained to adapt to the load, the material as a whole will become more resilient.

Indeed, materials based on disordered networks appear to be prime candidates for trainability. Many materials and patterns can be profitably modeled as networks. Macroscopic mechanical metamaterials composed of nodes and struts [29-31], crosslinked polymers [32-34] and biological fibers [35, 36], the abovementioned bone [1, 37], and even the creases in folded sheets [38] all have a network structure where links between nodes can be clearly identified. We can then think of training as evolving the properties of the links and/or the configuration of the nodes. Thus, networks can be viewed as *prototypical adaptive materials*. This adaptation can also involve hierarchies of networks [31]. The initial experiments for directed aging relied on a macroscopic, disordered network of struts connecting nodes, all cut from polymeric material [7]; the aging then changed the mechanical properties of the struts (while not altering the connectivity at the nodes) by reorganizing the more microscopic networks formed by the crosslinked polymer comprising the strut and node material. It has also been shown in computer simulations, that interactions between distant pairs of the macroscopic network nodes can be induced by aging the system while driving the nodes to have the desired response [39]. This again is a feature of networks that mimic the functions found in biological matter – in this case the allosteric properties of proteins [40-43]. In additional to real-space networks, training and memory have been shown to be relevant in networks of chemical affinities between many species of molecules [44-46].

One aspect of being in a network of fixed topology is that the connectivity of each of the nodes remains unaltered as the training proceeds. In this case, the unchanging topology encodes a memory of the initial configuration. The question that then presents itself is whether materials that are not representable as networks or have networks with evolving topology can be trained in the same manner as those that have a fixed network topology. That is, *can the memory of prior training – and its usefulness in creating novel function – be preserved if the material connectivity is allowed to change significantly over time?*

In answering this question we first consider a process that combines a training step with a separate step that involves the topology change. Such process evolves material function in ways that resemble modes of biological evolution through a combination of Darwinian and Lamarckian mechanisms. If we associate a given network topology with a genotype, the above discussion of in-situ training can be viewed as within-a-lifetime adaptation, while changes in network topology would be akin to genetic changes over generations in Darwinian evolution. The basic Darwinian mechanism might start with networks of slightly different topology and train them for a particular task or set of tasks. The most promising networks, in the sense that they respond best to training, become the initial `parent' network topology from which copies are made, potentially with local variations in topology. The variations could be due to 'mutations' in which some bonds are incorrectly copied to the next generation. These 'offspring' networks are then trained again for different tasks and the topologies that were most successful in being trained are again copied - with some variation - and trained again. Such a process resembles Darwinian evolution of networks but with selection on the ability to undergo successful training, rather than the usual optimization of materials for a specific task. The materials resulting from such selection do not promise to have any specific property but instead, can be expected to undergo successful training for a range of tasks.

During the training of each generation the local network geometry adapts as the struts deform from the applied strains. Importantly, in the copying step to generate the next generation, these deformations are copied too, creating an untrained network with the exact geometry (not only the topology) of the productively trained parent. Since the offspring network uses fresh material, it can be trained again to the fullest extent (in terms of parameters such as the stiffnesses). One can repeat the *in situ* training process from this fresh sample, make a copy of successfully trained networks (with variation) and repeat this over generations of material.

This type of training is analogous to Darwinian evolution with Larmackian elements, since the trained geometry of a network in one generation is inherited by the subsequent generation. Such a training scheme might expand the universe of trainable tasks; tasks that could not be trained for with one round of training might be achievable by refreshing the trained geometry. Another distinct Larmackian possibility is directed mutagenesis [47], where the training creates a set of bonds that are distorted and are therefore hard to reproduce from one generation to the next. The training thus promotes mutations at the spots where training is most effective. Schematically, we suggest: (training/aging) $\Rightarrow$ (shape of bonds at particularly important parts of the network) $\Rightarrow$ (makes those bonds difficult to copy correctly) $\Rightarrow$ (creates mutations – copying errors – preferentially at those spots) $\Rightarrow$ (pass to the next generation those changes originally caused by the training). This is an example mechanism in which training can influence the Darwinian evolution of the material.

More generally, the training and the topology change need not be separated cleanly. Thus training with concurrent topology changes can work in network-based materials as long as training memory can be preserved, meaning that relaxation back to an untrained state is sufficiently slow. For example, the stress-adaptive behavior in bone is a consequence of training that triggers changes in local network topology, and such bone remodeling can grow new connections where needed [1, 37]. Similar remodeling occurs in many other bio-mechanical networks [48]. Different examples are dry granular materials [49-51] or dense, non-Brownian suspensions of small particles in a liquid [52-55], where training by cyclic shear evolves the network of particle-particle contacts, thereby changing the local connectivity and instilling a memory of the pathway created by the applied shear strain. At the molecular scale, networks formed by dynamic covalent bonds among polymers can similarly evolve and adapt [34, 56, 57]. Examples here are new materials based on liquid crystal elastomers that can exhibit trainable shape shifting. These materials have many stable states at normal operating conditions and can be reconfigured because the covalent bonds can change neighbors: subjecting these materials to a new configuration during training induces dynamic bond exchange, which relaxes internal stresses and keeps a trained-in configuration stable.

Given that trainability requires the ability of a material to reconfigure internally, that same ability to change also implies that trained materials might be prone to 'forget' unless training stimuli continue to be applied [8]. However, as long as the material stays exposed to the intended operating conditions such 'refresher training' happens automatically through exposure to the environment. We experience the same with exercising our bodies. Still, depending on the strength of the training memory, there are at least two classes of trainable materials. The first and certainly more common class has a training memory that is finite and will decay after some time. The other class has a very long, potentially infinite memory due to a response that is strongly hysteretic between when a training stimulus is applied and when it is withdrawn. Examples are networks of coupled mechanical elements that each can reversibly snap into one of two states (so-called hysterons) [58].

In both cases, this brings up the question of how to devise protocols that efficiently erase previous training outcomes so that a material subsequently can be retrained for an entirely different response. This can be seen as another, but related, research area that studies not only the elastic response of a material but also the time dependence of the response – both for learning and forgetting as well as determining the pathway followed by the dynamics [59]. It also suggests that for adequate retraining, modifications of a material during the initial training must be reversible. This brings in the role not only of raising or lowering temperature as a way of enhancing or halting relaxation, but also the role that different chemical reactions can have in altering reactions to sections of a material that are under greater or lesser stress. This, again, takes inspiration from how biological material can alter its properties by sensing target bonds that are under large or small stress [60].

Trainability, together with learning and memory, forms key elements in the evolution of biological systems and, more recently, in machine learning with computers. Currently no systematic framework exists for designing trainable (soft or hard) materials and for devising optimal training protocols. One can hope to develop such a framework by taking advantage of an emerging synergy between recent ideas from biology, materials science, polymer chemistry and soft matter physics, all addressing different aspects of learning and memory in complex systems. These include theories of plasticity, evolution and learning in biology with recent insights about the malleability of disorder in material science. The new types of trainable materials we envision can exist from the macroscopic to the molecular scale; in their most potent manifestations they will require careful consideration of all these scales. The ultimate aim is to develop novel strategies for creating soft materials with multiple, targeted and adaptable functions. Research along this direction has the potential to lay the scientific foundation for training as a new paradigm within the larger field of materials processing. In this vision, we can use training to make matter functional in novel ways. Training will make matter matter in ways we have not yet discovered or even conceived.

**Acknowledgements:** We thank our many collaborators who worked on different aspects of training in materials. This work was primarily supported by the National Science Foundation

MRSEC under Award No. DMR-2011854. S.R.N. acknowledges support from DOE Basic Energy Sciences Grant No. DE-SC0020972.

**Competing Interest:** The authors declare no competing interest.